# Understanding Student Computational Thinking with Computational Modeling

John M. Aiken[*], Marcos D. Caballero[†], Scott S. Douglas[**], John B. Burk[††],
Erin M. Scanlon[**], Brian D. Thoms[*], Michael F. Schatz[**]

[*]Department of Physics and Astronomy, Georgia State University, Atlanta, GA 30303
[†]Department of Physics, University of Colorado Boulder, Boulder, CO 80309
[**]School of Physics, Georgia Institute of Technology, Atlanta, GA 30332
[††]Science Department, St. Andrew's School, Middleton, DE 19709

**Abstract.** Recently, the National Research Council's framework for next generation science standards highlighted "computational thinking" as one of its "fundamental practices". 9[th] Grade students taking a physics course that employed the Modeling Instruction curriculum were taught to construct computational models of physical systems. Student computational thinking was assessed using a proctored programming assignment, written essay, and a series of think-aloud interviews, where the students produced and discussed a computational model of a baseball in motion via a high-level programming environment (VPython). Roughly a third of the students in the study were successful in completing the programming assignment. Student success on this assessment was tied to how students synthesized their knowledge of physics and computation. On the essay and interview assessments, students displayed unique views of the relationship between force and motion; those who spoke of this relationship in causal (rather than observational) terms tended to have more success in the programming exercise.

**Keywords:** Physics education research, Computational Modeling, Computational Thinking, Modeling Instruction
**PACS:** 01.30.Cc

## INTRODUCTION

Numerical computation has fundamentally changed the way scientific research is done. Science relies more and more on models that require numerical computation to probe, so students must learn to extend their knowledge to include the use of numerical computation. Unfortunately, most high school students today are never introduced to computation's problem-solving powers. The lack of computation in domain-specific STEM courses is not addressed in most high school computer science courses, which typically focus on programming and procedural abstractions rather than solving science problems. In recognition of these shortcomings, the recently published National Research Council's (NRC) framework for next-generation K-12 science standards lists "computational thinking" as one of the fundamental practices that should be incorporated into future K-12 science curricula [1]. This presents a shift in the educational paradigm for students who are learning science. To move toward these standards, students must begin to engage in the practice of computational thinking, which in physics includes developing models of physical phenomena and learning to use a computer to solve, simulate, or visualize physical problems.

In this paper, we discuss briefly how we have integrated computation into an existing 9[th]-grade physics curriculum, present the results from three assessments of computational thinking, and close with our reflections as well as a discussion of future research directions.

## COMPUTATIONAL INSTRUCTION IN HIGH SCHOOL PHYSICS

We have worked with an in-service high school physics teacher for the past two years to develop a computational curriculum for a 9[th]-grade conceptual physics course. The high school instructor has used the Modeling Instruction physics curriculum [2] for several years. He has also presented simulations of physical phenomenon that were written using the VPython programming environment. VPython allows students to create three-dimensional simulations easily and to accompany those simulations with graphs and motion diagrams that update in real-time [3]. To facilitate instruction in computation, we have developed a suite of computational assignments (using VPython) that complement and enhance Modeling Instruction's treatment of force and motion topics [4].

During the fall semester, students developed computational models of four Modeling Instruction force and motion models (constant velocity, constant acceleration, balanced forces, and unbalanced forces) to predict the motion of objects described by various mathematical models (e.g., linear, quadratic). We confined our computational exercises to these four models (described by Newton's $2^{nd}$ Law) because computational modeling highlights the similarities between them [4]. In all computational activities, students used Euler-Cromer numerical integration [5] to determine the velocity and position after each time step. Students were also instructed to use the net force divided by mass in their program rather than simply the acceleration (e.g., baseball.v = baseball.v + Fnet/baseball.m * deltaT) to update the velocity. This emphasized the force's relationship to the equations of motion.

Computational assignments followed in-class experiments and problem-solving sessions. For example, while exploring the constant velocity model, students obtained and graphed data from wind-up cars. Students then constructed a computational model of a constant-velocity car. Students used these computational models to reproduce their experimental data and, later, to make predictions for a variety of physical situations to which the model applied.

## METHODS AND RESULTS

We implemented computational instruction in two separate $9^{th}$-grade physics classrooms with a total of 32 students. Each student had access to VPython on a laptop. Students also used the Georgia Tech-developed Python module PhysUtil [6]. PhysUtil was designed specifically to support the Modeling Instruction curriculum, and allows students to create graphs, motion diagrams, axes, and timers by writing only one or two lines of code.

### Assessment

Students' use of computation was evaluated with three separate assessments. Firstly, students attempted to develop a computational model of a physics problem using VPython in a proctored environment. Through this proctored assignment, we assessed whether students were capable of writing a VPython program without any aid. Success in this matter alone does not necessarily constitute success in modeling the physical system; students can write syntactically correct programs with incorrect physics. Analysis of students' code provided a cursory view of the types of challenges (whether syntactical or physical) the students faced when constructing a computational model. While it is important for students to write programs correctly, programming is not computational thinking [7]. To probe their reasoning, students were asked to complete a second assessment by answering an essay question designed to ascertain how they connected their computational model to the physics that the model described. In particular, they were asked to describe how their computational model related to the physical model via the iterative loop. Analysis of the essay responses indicated that we needed to delve more deeply into student reasoning. Therefore, a subset of five students was selected to participate in a final think-aloud interview in which they described how to develop a computational model for a particular physical phenomenon. To provide a representative sample of students, we selected participants from a cross-section of different performance levels on the previous assignments.

### Proctored Assignment

For the proctored assignment, students attempted to develop a 2D computational model that determined the location and velocity of a thrown baseball after a specified amount of time. Students completed this model individually and without aid from their instructor. The proctored assignment was deployed on an online homework system. Students were provided with a program scaffold that imported the necessary modules, created the objects (baseball and ground), and defined the integration loop structure. To complete the assignment successfully, students would assign the appropriate initial conditions and complete the integration loop by employing Euler-Cromer integration [5]. To facilitate students' successful completion of this assignment, students were given a "Code Checking Case" [8]. In the Code Checking Case, students were provided with the correct final position and velocity of the ball after the given time had elapsed. Students could use this case to check if their program modeled the situation correctly. After completing the Code Checking Case, students modeled a similar physical situation for the "Grading Case". In the Grading Case, the initial conditions were altered (including the integration time) and the system was moved from the Earth to the surface of the moon (reduced gravity). Answers were not provided for the Grading Case. Students input their final answers (baseball's final location and velocity) and uploaded their code to the homework system.

We sought to determine students' success rates and if their struggles were due to challenges with physics or with computational modeling. Our analysis of student code suggests that high school students can engage in computational thinking in the context of physics and that these students are generally capable of using numerical computation to solve physics problems. Figure 1 summarizes our findings.

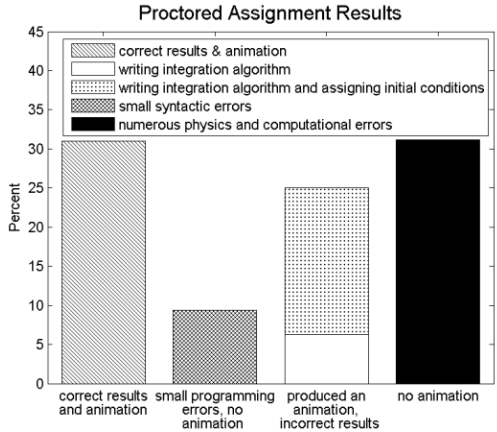

**FIGURE 1.** Students fell into four distinct groups. Less than half of these students struggled with programming errors (N=32). The distinction between animation and no animation is important because the animation is part of the computational representation the student is creating.

*Essay Question*

The code that the students wrote for the proctored assignment demonstrated a variety of output, but an assessment of the output alone was unable to probe deeply how students constructed these computational models. Students responded to an essay question after they completed the proctored assignment. Only 29 of 32 students completed the essay question. Students could run a working version of the program before answering the essay question. The essay question investigated whether students' success was predicated on simply reproducing an algorithm, or whether successful students made deeper connections between the physics and the computational algorithm. That is, did these students engage in the practice of computational thinking while developing their computational model?

The practice of computational thinking requires a logical problem solving approach that often involves thinking iteratively [8]. To further investigate how students developed their computational models, we asked students to describe the integration loop mathematically, physically, and programmatically. In order to provide a complete explanation, students needed to comment on the iterative procedure of the loop itself and its relationship to the integration of the equations of motion by the incremental stepping of Newton's Second Law.

The explanations presented by students in response to this question were captured by four distinct but not necessarily exclusive views. Some (38%) students presented a "force-causal" view of the loop structure. This view was characterized by a clear connection between force and motion. A student presenting a force-causal view would describe how the force of gravity would change the motion of the ball; "The loop is constantly changing the velocity of the ball while the [net force] stays constant. It makes the ball fall faster with every loop that runs". Another group (17%) of students presented a "kinematic-observational" view of the loop structure. These students indicated they had observed an acceleration (or some change in a kinematic quantity), but these students did not connect this observation back to the concept of a non-zero net force. One student with a kinematic-observational view noted, "The loop shows the changes in every vector as the time changes." Almost two-thirds (65%) of students described the integration loop as a local, iterative process governed by instantaneous influences. This iterative-local view was characterized by a discussion of incremental steps through the loop and statements such as "in this program, the [integration] loop is what the computer runs through to [compute] a new position, velocity, and all other forces for every [time it executes]." All the students who exhibited a force-causal view and nearly all students who presented a kinematic-observational view of motion also exhibited an iterative-local view of motion. Slightly more than a quarter (28%) of all respondents fell into no category. This group of students most often wrote very short, incomplete responses that were too difficult to accurately classify.

We compared the views that students presented on the essay question to their performance on proctored coding assignment. Students with each view were binned into the broad proctored assignment categories (i.e., "correct results and animation", "produced animation, but incorrect results", and "produced no animation"). Students who presented both an iterative-local and force-causal view were most likely to produce a correct program. Students whose essay were short and incomplete were most likely to write programs that produced no animations. Figure 2 summarizes our findings.

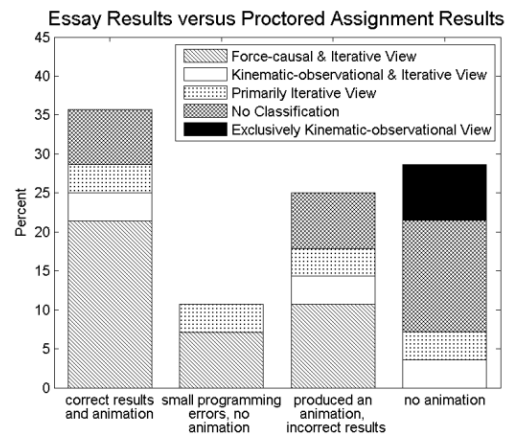

**FIGURE 2.** Students who displayed Force-casual and Iterative views were more likely to be successful on the proctored assignment (N=29).

*Interview and Think-Aloud Study*

Students' essay responses revealed that the concepts of force, motion, and iterative processes should be connected during instruction to facilitate computational thinking. However, investigating how students make these connections requires observing and questioning students while they engage in the practice of computational thinking. Several weeks after students completed the essay question, we interviewed them while they filled in the missing pieces of a scaffolded computational modeling program on paper. During the interview, students also answered questions about how they define a force and how forces, motion, and the integration loop are related. Students were asked to speak out loud while completing the scaffolded code and answering questions; their responses were videotaped. Only students whose proctored assignment code produced animations (i.e., "correct results and animation" and "produced animation, but incorrect results") were invited to the study. Six students were chosen to participate; five completed the interview. Of the students who completed the interview, 3 presented force-causal and iterative-local views on the essay question. One student had previously presented both a kinematic-observational and an iterative-local view, but expressed a force-causal and an iterative-local view in the interview. The last student presented a primarily iterative-local view on the essay question and in the interview.

For students who developed a correct computational model, the interviews further highlighted the links they made between force, motion, and iterative processes. A student who wrote a correct program described her code with a force-causal and an iterative-local view, "To predict the velocity you would have to do baseball.v = initial velocity of the baseball plus gravity times time. *That would give me the new velocity after* [the execution of] *every single loop.* And then you need to update the position based on the loop." This student mentions the basic concepts behind Newton's 2$^{nd}$ law but also describes how the numerical integration loop updates the velocity of the ball in each execution. By contrast, another student who constructed a model that produced incorrect animation demonstrated an incorrect conception of force and motion, "*force generally [is] acquired through motion.* There's always force acting on an object." When questioned about how the loop models the physics of the system, the student presented solely an iterative-local view, "[the loop] has formulas that it solves for, like, update position equals [baseball.pos + baseball.v*deltat]." While this student was able to generate a computational model for the proctored assignment that ran without (syntactic) errors, she did not use the correct physics to do so.

## DISCUSSION & IMPLICATIONS

Students in a 9$^{th}$-grade Modeling Instruction physics course were introduced to numerical computation as a means of predicting the motion of a physical system. After instruction, roughly a third of students were able to successfully complete an individual assessment in which they constructed a model of a new physical system. Student success on the proctored assignment was closely tied to how students synthesize knowledge of physics (force and motion) and computation (iterative processes). By contrast, students who described iterative processes but had not yet connected the concepts of force and motion were unable to create precise computational models. Future work aims to expand the data pool to more precisely characterize student views.

By using instantaneous influences on the object to describe the motion, successful students constructed a "model" of the physical system that employed a series of local "rules" to predict the motion. Learning to employ this model leads to a relatively robust problem-solving strategy. To solve new problems, only the "rule" for the net force needs to be changed. Moving forward, it becomes important to understand how students transfer this "model" to other problems.

## ACKNOWLEDGEMENTS

The authors would like to acknowledge the National Science Foundation's support of this research (Awards: DUE0618519 and DUE0942076).